\documentclass[twocolumn, prl, showpacs]{revtex4}

\usepackage{epsfig}

\usepackage{amsmath}
\usepackage{amsfonts}
\usepackage{amssymb}

\newcommand{\bra}[1]{\langle #1|}
\newcommand{\ket}[1]{|#1\rangle}

\newcommand{\vectornorm}[1]{\left|\left|#1\right|\right|}

\begin{document}

\title{Spectral coarse-graining of complex networks}

\author{David Gfeller} 
\author{Paolo De Los Rios}
\affiliation{Laboratoire de Biophysique Statistique, SB/ITP ,\\Ecole Polytechnique F\'ed\'erale de Lausanne (EPFL), CH-1015, Lausanne, Switzerland}

\date{\today} 

\begin{abstract}

Reducing the complexity of large systems described as complex networks is key to understand them and a crucial issue is to know which properties of the initial system are preserved in the reduced one. Here we use random walks to design a coarse-graining scheme for complex networks. By construction the coarse-graining preserves the slow modes of the walk, while reducing significantly the size and the complexity of the network. In this sense our coarse-graining allows to approximate large networks by smaller ones, keeping most of their relevant spectral properties.
 
\end{abstract}

\pacs{89.75.Fb, 02.50.-r}

\maketitle

One of the most difficult hurdles in the analysis and visualization of large complex networks \cite{Albert2002, Newman2003-2, Dorogovstev2002, Boccaletti2006} is, understandably, their sheer size. Given that most algorithms used to extract information from a network
topology run in times that grow polynomially with the number $N$ of network nodes, even networks of a few thousands nodes can represent a challenge, and networks with $N > 10^5$ become almost impossible to deal with.

A promising way around this problem is to \textit{coarse-grain} the network, i.e. to reduce 
the number of nodes and edges by means of a mapping of the network with $N$ nodes and $E$ edges into a smaller network with $\tilde{N}$ nodes and $\tilde{E}$ edges.
$\tilde{N}$ and $\tilde{E}$ have to be small enough to be amenable to analysis and visualization.

Several coarse-graining schemes have been proposed in the literature. 
The $k$-core decomposition, which is a node decimation approach, was first proposed in \cite{Bollobas1984} to isolate the central core of a network, and was shown to be extremely effective for visualization purposes \cite{Alvarez-Hamelin2005}. Alternatively, the number of nodes can be reduced by clumping them together in clusters. 
A widely accepted technique is based on community detection \cite{Girvan2002}. Within this framework, groups of nodes with more edges pointing to each other then to the rest of the network are considered as one single unit. After grouping the nodes a much reduced ``network of clusters'' is obtained, representing the functional units of a network.
Because of the importance and the complexity of finding meaningful clusters, a very large number of clustering algorithms have been developed recently \cite{Girvan2002, Reichardt2004, Donetti2004, Palla2005, Guimera2005-1, Newman2006-1, Capocci2004}. However there is often no clear statement on whether properties of the initial network are preserved in the network of clusters.

In this respect, the box-covering  
technique recently introduced by  Song {\em et al.}~\cite{Song2005} and further analyzed by Goh {\em et al.} \cite{Goh2006} deserves
a special mention: after covering \textit{self-similar} networks with suitably defined boxes of a given size, the new networks obtained by substituting each box with a node preserve some of the topological features of the original ones. Song {\em et al.} have thus recognized that network reduction should go hand-in-hand with the preservation of some relevant network properties, akin to the
renormalization group in statistical physics. What properties should be preserved is instead an unsettled issue: although the network topology can provide important clues about the organization of the system under scrutiny, it does not necessarily bear any insight on the internal dynamics of the system. 

In this Letter we introduce a mathematical framework to coarse-grain networks, based on the idea of grouping nodes together.  Thus our starting point is similar to clustering approaches. However, contrary to the clustering paradigm of identifying the ``correct'' communities in a network, our goal is to obtain a reduced network that preserves some properties of the initial one. Here the properties of interest will be the main characteristics of random walks on networks~\cite{Noh2004}. 

Random walks play a central role in a large number of dynamical processes taking place on complex networks.
Their evolution is described by a stochastic matrix $W$. If $A$ is the adjacency matrix, then $W_{ij}=A_{ij}\left(\sum_lA_{lj}\right)^{-1}$ gives the transition probability from $j$ to $i$. $W$ has several interesting properties. In particular for a connected undirected network, the Perron-Frobenius theorem states that the largest eigenvalue $\lambda^1$ is equal to 1.
The right eigenvector $\ket{p^1}$ associated to $\lambda^1$ describes the stationary state and the left eigenvector $\bra{u^1}$ is constant.
Typically, eigenvectors with eigenvalues close to one capture the large-scale behavior of the random walk. % whereas eigenvectors with smaller eigenvalues contain the small-scale behavior.
Since a coarse-grained network necessarily loses the fine details of the original one, expectedly our goal is to preserve the large scale behavior of the random walk, hence its largest eigenvectors and eigenvalues.

As a starting point, we want to ensure that two nodes in an undirected network (say nodes 1 and 2) having exactly the same neighbors are grouped together, since they cannot be distinguished from the point of view of a random walk. In terms of a left eigenvector $\bra{u^{\alpha}}$ of $W$ it means that $u^{\alpha}_1=u^{\alpha}_2$ for any $\lambda^\alpha \neq 0$ since column 1 and 2 of $W$ are equal.
The obvious coarse-graining step is to coalesce such pair of nodes, with the resulting new node carrying the
sum of the edges of the initial ones. The new network in which node 1 and 2 have been merged is characterized by a $(N-1)\times (N-1)$ adjacency matrix $\tilde{A}$, with the first line, resp. column, of $\tilde{A}$ being the sum of the two first lines, resp. columns, of $A$. On this reduced network, the stochastic matrix $\tilde{W}$ describing a random walk is obtained by normalizing the columns of $\tilde{A}$.

At this stage, it will be useful to write $\tilde{W}$ as a product of three matrices,
\[\tilde{W}=RWK.\]
$K$ and $R$ are two projection-like operators from the $N$-dimensional space of the initial nodes to the ($N$-1)-dimensional space of the new nodes. In order to fulfill the definition of $\tilde{W}$ and using that $p^1_j\propto \sum_iA_{ij}$ for undirected networks, $K$ and $R$ are defined as :

\[
K=\left(\begin{array}{cc}
\frac{p^1_1}{p^1_1+p^1_2} & 0  \hdots 0\\
\frac{p^1_2}{p^1_1+p^1_2} & 0 \hdots 0\\
\begin{array}{c} 0\\ \vdots \\ 0\end{array} & I_{N-2}
\end{array}
\right)
\hspace{2mm}\text{and}\hspace{2mm}
R=\left(\begin{array}{cc}
\begin{array}{cc} 1 & 1\end{array} & 0 \hdots 0\\
\begin{array}{cc} 0 & 0 \\ \vdots &\vdots \\ 0 & 0\end{array} & I_{N-2}\\
\end{array}
\right),
\]
(see \cite{Lafon2006} for a similar mathematical framework). The interesting features of $\tilde{W}$ come from the property~\cite{Meila2001} that {\em if} $u^{\alpha}_1=u^{\alpha}_2$, {\em the vector} $\bra{u^{\alpha}}K$  {\em is a left eigenvector  of} $\tilde{W}$ {\em with eigenvalue} $\lambda^{\alpha}$ (i.e. $\bra{u^{\alpha}}K=\bra{\tilde{u}^{\alpha}}$).
To obtain this result one simply needs to see that $\bra{u^{\alpha}}KR=\bra{u^{\alpha}}$ if $u^{\alpha}_1=u^{\alpha}_2$. 
In the case of undirected networks, the result can be extended to the right eigenvectors.
Under the same hypothesis ($u^{\alpha}_1=u^{\alpha}_2$), {\em the vector} $R\ket{p^{\alpha}}$  {\em is a right eigenvector of} $\tilde{W}$ {\em with eigenvalue} $\lambda^{\alpha}$ (for $\alpha=1$ the result holds as well in directed networks).
Moreover we could show analytically that a perturbative approach can be carried out. If $|u^{\alpha}_1-u^{\alpha}_2|\propto \epsilon$ for a given $\lambda^\alpha\neq 0$, $\bra{u^{\alpha}}K$, $R\ket{p^{\alpha}}$, resp. $\lambda^{\alpha}$, need to be corrected by vectors, resp. a scalar, scaling as $\epsilon$ in order to become left and right eigenvectors, resp. eigenvalue, of $\tilde{W}$.
%Numerical computations indicate that the perturbative approach is robust (see the examples below).

To summarize we have introduced a mathematical framework such that grouping nodes with similar components in $\bra{u^{\alpha}}$ has a spectral interpretation: it preserves the eigenvalue $\lambda^\alpha$, averages the components of $\bra{u^{\alpha}}$ and for undirected networks sums up the components of $\ket{p^{\alpha}}$.

For simplicity the case where only two components of an eigenvector are equal (resp. close to each other) has been considered. It is straightforward to generalize the grouping to all nodes having the same components (resp. components close to each other) in $\bra{u^{\alpha}}$. Groups are first labeled from 1 to $\tilde{N}$ and $\delta_{C,i}$ is defined as $1$ if node $i$ belongs to group $C$, $0$ otherwise ($C=1\hdots \tilde{N}$). Then $K$ and $R$ read $R_{Ci}=\delta_{C,i}$ and $K_{iC}=\frac{p^1_i}{\sum_{l\in C} p^1_l}\delta_{C,i}$, with $R$ a $\tilde{N}\times N$ matrix and $K$ a $N\times \tilde{N}$ matrix.

The method can be further extended to more than one eigenvector and groups are defined as the nodes having the same components (resp. components close to each other) over a set of left eigenvectors $\left\{\bra{u^{\alpha_i}}\right\}_{i=1}^S$. Choosing the $S$ first non-trivial eigenvectors ensures to conserve the slow modes of the random walk.

Spectral properties of $W$ have been used previously in spectral clustering techniques~\cite{Capocci2004}. However the properties derived above show that there exists a way to preserve the spectral properties of a network while reducing its size, which is the ultimate goal of any coarse-graining strategy.

\begin{figure}[]
\begin{center}
\includegraphics[width=55mm]{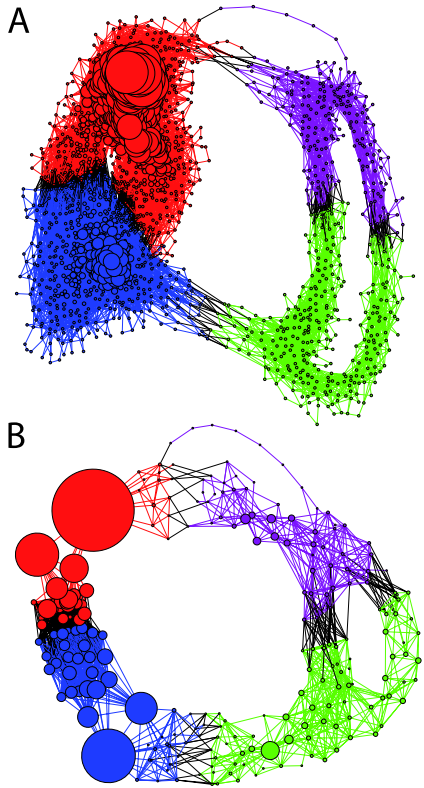}
  \caption{(Color online) A: Di-alanine folding network ($N$=1832). Node size is proportional to their weight (i.e. the number of times nodes have been visited in the simulation). The four different colors correspond to the clusters found in \cite{Gfeller2007-1}. B: Coarse-grained network ($\tilde{N}=227$) according to $\bra{u^{\alpha}}$, $\alpha=2,3,4$. Node size is proportional to the total weights of the groups. Colors correspond to the clusters in which the nodes of each group have been classified.}
  \label{ala}
\end{center}
\end{figure}

To illustrate our coarse-graining scheme, we applied it to the di-alanine folding network studied in \cite{Gfeller2007-1}, considered as undirected. The network was built from Molecular Dynamics simulation of a di-alanine peptide and consists of 1832 nodes (Fig.~\ref{ala}A). A node in Fig.~\ref{ala}A accounts for a configuration sampled during the simulation and edges represent transitions between configurations~\cite{Rao2004}. The weight of an edge between two configurations is equal to the total number of transitions sampled during the simulation. In a previous work~\cite{Gfeller2007-1}, the di-alanine folding network was shown to consist of four main clusters (colors in Fig.~\ref{ala}), corresponding to the four main energy basins of the underlying free-energy landscape.
Random walks on such networks are representative of peptide dynamics since the elements of $W$ correspond to the effective transition probabilities, as observed along the simulation.

To coarse-grain the network, we have used the first three non-trivial left eigenvectors $\bra{u^2}, \bra{u^3}$ and $\bra{u^4}$ of $W$. Along each eigenvector, $I=60$ intervals of equal size have been defined between the highest and the lowest component. Nodes have been grouped together if they belonged to the same interval along the three eigenvectors. In this way 227 non-empty groups have been found. The coarse-grained network is shown in Fig.~\ref{ala}B. Colors were set according to the clusters of the nodes in each group. Clearly the coarse-grained network is not equivalent to the network of clusters. Although the nodes of a group do not necessarily belong to the same cluster, this situation happened only for 4 groups (representing 15 nodes) out the 227. We also applied on the coarse-grained network the same clustering algorithm \cite{Enright2002} used to identify the clusters in Fig.~\ref{ala}A. Exactly 4 clusters were obtained corresponding to more than 98\% of the initial nodes correctly classified. Thus, even if the aim of our coarse-graining approach is different than the usual clustering, the results are indeed consistent with the global features revealed by the cluster structure of the network. Moreover the cluster structure is robust under coarse-graining.

\begin{table}
\begin{center}
\begin{tabular}{|c|c|c|c||c|c|}
  \hline
   & $\alpha$ & $\lambda^\alpha$ & $\tilde{\lambda}^\alpha$ &  $\frac{\bra{u^\alpha}K\ket{\tilde{u}^\alpha}}{\vectornorm{\bra{u^\alpha}K}\cdot||\tilde{u}^\alpha||}$&  $\frac{\bra{\tilde{p}^\alpha}R\ket{p^\alpha}}{\vectornorm{R\ket{p^\alpha}}\cdot||\tilde{p}^\alpha||}$\\
  \hline
  %1 & 1 & 1\\
  {\bf A} & 2 & 0.99987 & 0.99987 & 0.9999 & 0.9999\\
          & 3 & 0.99947 & 0.99944 & 0.9998 & 0.9999\\
          & 4 & 0.99785 & 0.99780 & 0.9999 & 0.9999\\
  \hline\hline
 {\bf B} & 2 & 0.98955 & 0.98922 & 0.9985 & 0.9941\\
   & 3 & 0.98901 & 0.98861 & 0.9989 & 0.9924\\
   & 4 & 0.98779 & 0.98741 & 0.9885 & 0.9686\\
  \hline\hline
  {\bf C} & 2 & 0.99971 & 0.99971 & 0.999916 & 0.9999\\
   & 3 & 0.99934 & 0.99933 & 0.9994 & 0.9988\\
   & 4 & 0.99917 & 0.99916 & 0.9998 & 0.9997\\
  \hline
\end{tabular}
\caption{Columns~2 and~3: the three largest (non-trivial) eigenvalues of the stochastic matrices $W$ and $\tilde{W}$. Column~4: Scalar product between $\bra{u^\alpha}K$ and $\bra{\tilde{u}^\alpha}$ for the three left eigenvectors used in the coarse-graining procedure. Column~5: Scalar product between $R\ket{p^\alpha}$ and $\ket{\tilde{p}^\alpha}$ for the three right eigenvectors. Box {\bf A}: Di-alanine network shown in Fig.~\ref{ala}A and B. Box~{\bf B}: Erd\"os-R\'enyi network. Box~{\bf C}: Barab\'asi-Albert network.}
\label{table}
\end{center}
\end{table}

As expected from our perturbative derivation, the first eigenvalues are preserved in the coarse-grained network with high accuracy (Table \ref{table} {\bf A} columns 2 and 3). Moreover the normalized scalar product (Table \ref{table} {\bf A} columns 4 and 5) shows that the projected eigenvectors $\bra{u^{\alpha}}K$ and $R\ket{p^{\alpha}}$ are almost equal the corresponding eigenvectors of $\tilde{W}$. Similar results have been obtained considering the giant component of an Erd\"os-R\'enyi network~\cite{Erdos1959} ($N=5626$, $<k>=2$, Table \ref{table} {\bf B}) and a Barab\'asi-Albert network~\cite{Barabasi1999} ($N=6005$, $m=1$, Table \ref{table} {\bf C}), always considering the three first non-trivial left eigenvectors $\bra{u^2}, \bra{u^3}$ and $\bra{u^4}$ and $I=60$.
The general agreement indicates that our perturbative approach is robust for various kinds of networks even if components in $\bra{u^{\alpha}}$ are not equal but close to each other within the groups.

\begin{figure}[]
\begin{center}
\includegraphics[width=85mm]{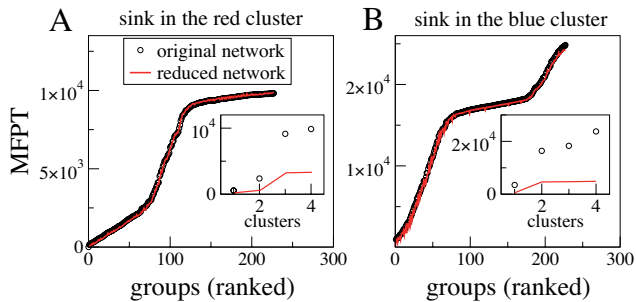}
  \caption{(Color online) Comparison of the MFPT. The circles ($\circ$) represent the average MFPT ranked for each group in the original network (variances are not shown since they are always smaller than the size of the circles). The MFPT of the corresponding nodes in the coarse-grained network is displayed with red lines. A: di-alanine folding network with the sink $i$ as the heaviest node of the red cluster . B: di-alanine folding network with the sink $i$ as the heaviest node of the blue cluster. Onsets: Comparison of the MFPT between the original network ($\circ$) and the network of clusters (red line).}
  \label{mfpt}
\end{center}
\end{figure}

Fig.~\ref{ala} hints that the global architecture of the coarse-grained network is representative of the original one. For instance most of the nodes buried in the center of the red cluster form one single group, while the nodes lying along the few paths connecting the red and violet clusters, and therefore critical for the network global connectivity, are well preserved. A more stringent test is done by comparing the mean first passage time (MFPT) from node $j$ to node $i$, $T_{ij}$.
In the context of transport phenomena or search on a network, MFPT is an important characteristic of random walks \cite{Noh2004, Baronchelli2006}. To compute it exactly, one usually considers node $i$ as a sink and uses the stochastic matrix $\hat{W}$ with the $i^{\text{th}}$ column set to 0 ($T_{ij}=\sum_{t=0}^{\infty}t(\hat{W}^t)_{ij}=\sum_{\alpha=1}^N\hat{u}^\alpha_i\hat{p}^\alpha_j\frac{\hat{\lambda}^\alpha}{(1-\hat{\lambda}^\alpha)^2}$). 
To compare the MFPTs, we used the coarse-graining shown in Fig.~\ref{ala}B, defining the sink node $i$ as a single group. Fig.~\ref{mfpt} shows with black circles ($\circ$) the average MFPT to node $i$ for each group in the original network. The MFPT to the group consisting of node $i$ in the coarse-grained network is displayed with red lines. The excellent overlap indicates that the MFPT is extremely well preserved, whereas this is not the case in the network of clusters (see onsets in Fig.~\ref{mfpt}). Hence the coarse-grained network is representative of the general features of the diffusion process in the initial network. Moreover this finding was shown to be robust if other eigenvectors are included, as long as the size of the intervals is kept small enough. In this respect the value $I\propto \epsilon^{-1}$ tunes the degree of precision: increasing $I$ improves the agreement between the initial and the coarse-grained network, but in the same time results in a larger $\tilde{N}$.

In general the large eigenvalues and eigenvectors of $W$ represent the large scale behavior of random walks. However, in some cases eigenvectors are directly associated with useful quantities.
This is the case with the PageRank matrix~\cite{Brin1998}.% PageRank has received much attention because of its use in search engines like Google. 
PageRank is defined via a stochastic process on the WWW, where for each step a ``random surfer'' either follows with probability $d$ one of the existing outgoing links or jumps at random to another site with probability $(1-d)$. The PageRank of the nodes corresponds to the stationary state of the process (i.e. to $\ket{p^1}$). Thus under coarse-graining the PageRank in the reduced network is the sum over the PageRank in the initial network. The only effect of the directed nature of the WWW, compared to undirected network, is that nothing ensures that $\ket{p^\alpha}$ is preserved when coarse-graining along $\bra{u}^\alpha$ for $\alpha >1$.

As a second example, we consider the exit probability on a network with two nodes (say node 1 and $N$) in which the random walk is trapped. 
One can show that the exit probability in node 1, resp. $N$, starting at $j$ can be expressed as the left eigenvector $\bra{u^1_j}$, resp. $\bra{u^2_j}$, of the stochastic matrix describing the transition probabilities and including the two traps.
Hence coarse-graining the network according to the eigenvector $\bra{u^1}$, resp. $\bra{u^2}$, is equivalent to preserving the exit probabilities in node 1, resp. $N$. 
In the case of a network describing the dynamics of a peptide, as the example studied in this Letter, the exit-probability can be associated with the p-fold \cite{Du1998}, defined as the probability to reach the native state before the denaturated state.
If the two traps are chosen as representatives of the native and denaturated states (for instance the heaviest nodes of the two main clusters of Fig. \ref{ala}), our method allows to coarse-grain the network in such a way that the p-fold is perfectly preserved for every configuration (see \cite{Rhee2005} for a related coarse-graining framework considering a continuous diffusion process).

In conclusion, we have defined a mathematical framework for coarse-graining complex networks based on random walks. This operation has the intrinsic property of preserving the first eigenvalues and the corresponding eigenvectors. In this sense it can be regarded as a decimation of the fast modes, without altering the slow modes, akin to $k$-space coarse-graining, and eventually coming back to a real space coarse-grained network. Moreover we have shown for a network on which random walks have a physical interpretation that the coarse-graining provides a highly representative approximation of the initial network, giving rise to a way to circumvent the large size of complex networks for their analysis and visualization. Finally, from a computational point of view, the first eigenvectors are fast to calculate with the existing optimized routines for sparse matrices. Therefore our method can be easily utilized on large networks.

We are thankful to David Morton De La Chappelle, Francesco Rao and Amedeo Caflisch for a critical reading of the manuscript. This work was financially supported by COSIN (FET Open IST 2001-33555), DELIS (FET Open 001907) and the SER-Bern (02.0234).

\end{document}